%% file: Burlak.tex
\documentclass[aps,preprint]{revtex4}%
\usepackage{amsfonts}
\usepackage{amsmath}
\usepackage{amssymb}
\usepackage{graphicx}%
\begin{document}
\title{The dynamics of coupled atom and field assisted by continuous external pumping.}
\author{Gennadiy N. Burlak}
\affiliation{Center for Research on Engineering and Applied Sciences, Autonomous State
University of Morelos,Cuernavaca, Mor. Mexico. Corresponding author: E-mail
address: gburlak@uaem.mx(G.N.Burlak)}

\begin{abstract}
The dynamics of a coupled system comprised of a two-level atom and cavity
field assisted by continuous external classical field (driving Jaynes-Cummings
model) is studied. When the initial field is prepared in a coherent state, the
dynamics strongly depends on the algebraic sum of both fields. If this a sum
is zero (the compensative case) in the system only the vacuum Rabi
oscillations occur. The results with the dissipation and external field
detuning from the cavity field are also discussed.

\end{abstract}
\maketitle

\section{Introduction}

The ability to create, manipulate, and characterize quantum states is becoming
an increasingly important area of physics research, with implications for such
areas of technology as quantum computing, quantum cryptography, and
communications, see \cite{Blinov:2004a}, \cite{WojciechHubertZurek:2003a},
\cite{Raimond:2001a}, \cite{Brune:1996a}. Most of research in quantum
nonlocality and quantum information is based on entanglement of two-level
particles. One of the most interesting aspects of its dynamics is the
entanglement between atom and field states. This essentially quantum
mechanical property with no classical analog is characterized by the
impossibility of completely specifying the state of the global system through
the complete knowledge of the individual subsystem's dynamics.

The Jaynes-Cummings model\cite{Jaynes:1963a} (JCM) for the interaction between
a two-state atom and a single mode of the electromagnetic field holds a
central place in description of such interaction and provides important
insight into the dynamical behavior of atom and quantized field. In driving
JCM the cavity field and driving field start to interact, which provide a
possibility to study directly the field dynamics at joint interaction with a
two-level atom. Recently, it was shown that the effective coupling between an
atom and a single cavity field mode in JCM (driving JCM) can be drastically
modified in the presence of a strong external driving field
\cite{Solano:2003a},\cite{Shi-BiaoZheng:2002a}, \cite{Shi-BiaoZheng:2003a},
\cite{Lougovski:2004a}. The important line of this direction is to use
microcavities and microspheres for changing the features of atom-field
interaction as a result of placing atom or quantum dots into a microcavity
(see\cite{Vahala:2003a}, \cite{Artemyev:2001a}, \cite{Burlak:2003a}).

The driven Jaynes-Cummings model for cases where the cavity and external
driving field are close to or in resonance with the atom, has been studied by
several authors. In Ref.\cite{Alsing:1992a} studied the Stark splittings in
the quasienergies of the dressed states resulting from the presence of the
driving field in the case where both fields are resonant with the atom.
Authors\cite{Jyotsna:1993a} studied the effect of the external field on the
Rabi oscillations in the case where the cavity field is resonant with the atom
and where the external field is both resonant and nonresonant. In
Ref.\cite{Dutra:1993a} studied a similar model where the external field was
taken to be quantized. Much attention was given to the limit of high-intensity
of driving field. In Ref.\cite{Chough:1996a} have studied the JCM with an
external resonant driving field and have shown that the collapses and revivals
of the mean photon number occur over a much longer time scale than the revival
time of the Rabi oscillations for the atomic inversion.
Author\cite{Gerry:2002a} studied the interaction of an atom with both a
quantized cavity field and an external classical driving field, in the regime
where the atom and fields are highly detuned. He has shown how dispersive
interaction can be used to generate coherent states of the cavity field and
various forms of superpositions of macroscopically distinct states.

The main goal of the present work is the calculation of the dynamics of
coupled atom and field assisted by continuous external pumping in case when
the initial field is prepared in a coherent state. The main result is the
following: Starting with a field's mode in a coherent state and with the atom
in its upper state, the dynamics strongly depends on the algebraic sum of
amplitudes of initial cavity field and the external field. If such sum is
close to a zero (the compensative case), in system only the vacuum Rabi
oscillations occur.

This Letter is organized as follows. In Section 2 we discuss the motion
equations for two-level atom coupled to the field in cavity with the
assistance of continue pumping classical field. Section 3 presents the results
of numerical study for the dynamics of the atom and field subsystems for
dissipative case by the technique of the master equation. The behavior of
entropy and Fourier spectrum of oscillations is studied also. In last Section,
we discuss and summarize our results.

\section{Basic equations}

Consider a two-state atom, driven by a classical external field $(1/2)E_{e}%
\exp(i\omega_{e}t)+c.c.$, and coupled to a cavity mode of the quantized
electromagnetic field. The Hamiltonian for the atom-cavity system (assuming
$\hbar=1$) in rotating-wave approximation (RWA) is given by%

\begin{equation}
H=\frac{1}{2}\omega_{0}\sigma_{3}+\omega_{a}a^{+}a+g\left[  \sigma^{-}%
a^{+}+\sigma^{+}a\right]  +\frac{1}{2}\left[  \mathcal{E}\sigma^{+}%
e^{i\omega_{e}t}+\mathcal{E}^{\ast}\sigma^{-}e^{-i\omega_{e}t}\right]
\text{,} \label{Hamil full dimless}%
\end{equation}
where $\omega_{0}$\ is the transition atom frequency, $\omega_{a}$ is the
cavity frequency, $g$ is the coupling constant between the atom and the cavity
field mode, $\mathcal{E}$ is proportional to the coupling constant between the
atom and the external classical field of frequency $\omega_{e}$ and the
amplitude of that field, $a^{+}$ and $a$ are the creation and annihilation
operators for the cavity mode $[a,a^{+}]=1$. In general $\omega_{0}$ ,
$\omega_{a}$, and $\omega_{e}$ are different. To remove the time dependence in
$H$, we use the operator $\exp[-i\omega_{e}t(\sigma_{3}+a^{+}a)]$ to transform
to a frame rotating at the frequency $\omega_{e}$. The Hamiltonian in the
rotating frame (the interaction picture) is then%

\begin{equation}
H_{i}=\frac{\Delta}{2}\sigma_{3}+g\left[  \sigma^{-}a^{+}e^{-i\delta t}%
+\sigma^{+}ae^{i\delta t}\right]  +\frac{1}{2}\left[  \mathcal{E}\sigma
^{+}+\mathcal{E}^{\ast}\sigma^{-}\right]  \text{,}
\label{full Hamil inter pict}%
\end{equation}
where $\Delta=\omega_{0}-\omega_{e}$, $\delta=\omega_{a}-\omega_{e}$,
$\sigma^{\pm}=\left(  \sigma^{x}\pm i\sigma^{y}\right)  /2$, $\sigma_{x}$,
$\sigma_{y}$, $\sigma_{z}$ are Pauli matrices. First we consider the resonant
case when $\Delta=0$ and $\delta=0$. Case of non-zero detuning $\delta\neq0$
is discussed in the second part. The resonant Hamiltonian ($\omega_{a}%
=\omega_{0}=\omega_{e}$) in the interaction picture has the form%

\begin{equation}
H_{ir}=g\left[  \sigma^{-}a^{+}+\sigma^{+}a\right]  +\frac{1}{2}\left[
\mathcal{E}\sigma^{+}+\mathcal{E}^{\ast}\sigma^{-}\right]  \text{.}
\label{Hamil inter dimless}%
\end{equation}
Further we use the following dimensionless variables $\tau=\omega_{0}t$,
$g/\omega_{0}\rightarrow g$, $\mathcal{E}/\omega_{0}\rightarrow\mathcal{E}$.
If $\mathcal{E}=0$ the Eq.(\ref{Hamil inter dimless}) describes the standard
JCM, case $\mathcal{E}\neq0$ corresponds to driving JCM. \ To obtain the
solution to Eq.(\ref{Hamil inter dimless})\ we introduce the displacement
operator $D(\gamma)=\exp\{\gamma a^{+}-\gamma^{\ast}a\}$, $\gamma
=\mathcal{E}/2g$, which allows us to rewrite Eq.(\ref{Hamil inter dimless}) in
the form%

\begin{equation}
H_{ir}=gD^{+}(\gamma)(\sigma^{+}a+a^{+}\sigma^{-})D(\gamma)\text{,}
\label{H before shift}%
\end{equation}
where identity \ $D^{+}(\gamma)(a^{+},a)D(\gamma)=(a^{+}+\gamma^{\ast
},a+\gamma)$\ is used. Establishing in (\ref{H before shift}) the Hamiltonian
$D(\gamma)H_{ir}D^{+}(\gamma)=g(\sigma^{+}a+a^{+}\sigma^{-})$ and the state
vector $\left\vert \widetilde{\psi}\right\rangle =D(\gamma)\left\vert
\psi\right\rangle $ we obtain the well-known the Schr\"{o}dinger equation of
the standard JCM in form%

\begin{equation}
i\frac{\partial\left\vert \widetilde{\psi}\right\rangle }{\partial\tau
}=(\sigma^{+}a+a^{+}\sigma^{-})\left\vert \widetilde{\psi}\right\rangle
\text{.} \label{Schr eq shifted}%
\end{equation}

Now consider the case when the initial state of the field in cavity is a
coherent state $\left\vert \alpha\right\rangle $, with $\alpha=\overline
{n}^{1/2}e^{-iv}$ ($\overline{n}$ is the average number of photons in the
field). Also we assume the atom is prepared in the excited state $\left\vert
e\right\rangle $ ($\left\vert g\right\rangle $ is the ground state). The
initial state vector $\left\vert \psi\right\rangle =\left\vert e\right\rangle
\left\vert \alpha\right\rangle =\left\vert e\right\rangle D(\alpha)\left\vert
0\right\rangle $ allows us to write for Eq.(\ref{Schr eq shifted}) the
corresponding initial state vector $\left\vert \widetilde{\psi}\right\rangle $ as%

\begin{equation}
\left\vert \widetilde{\psi}\right\rangle =D(\gamma)\left\vert \psi
\right\rangle =\left\vert e\right\rangle D(\gamma)D(\alpha)\left\vert
0\right\rangle =\left\vert e\right\rangle \left\vert \widetilde{\gamma
}\right\rangle \text{,}\ \label{init cond shifted}%
\end{equation}
where $\widetilde{\gamma}=\gamma+\alpha$ and\ overall factor $\exp
(i\operatorname{Im}(\gamma\alpha^{\ast}))$ is dropped. With
Eq.(\ref{init cond shifted}) the solution to the standard JCM
Eq.(\ref{Schr eq shifted}) is given by%

\begin{equation}
\left\vert \widetilde{\psi}\left(  \xi\right)  \right\rangle =%
{\displaystyle\sum\limits_{n=0}^{\infty}}
C_{n}(\widetilde{\gamma})\{\cos(\xi\sqrt{n+1})\left\vert e\right\rangle
\left\vert n\right\rangle -i\sin(\xi\sqrt{n+1})\left\vert g\right\rangle
\left\vert n+1\right\rangle \}\text{, }\xi=g\tau\text{,} \label{psi shifted}%
\end{equation}
where $C_{n}\equiv C_{n}(\widetilde{\gamma})=\exp(-\left\vert \widetilde
{\gamma}\right\vert ^{2}/2)\widetilde{\gamma}^{n}/\sqrt{n!}$ are expansion
coefficients for $\left\vert \widetilde{\gamma}\right\rangle $ state in the
number representation $\left\vert n\right\rangle $. The Eq. (\ref{psi shifted}%
) allows us to write the solution to resonant driving JCM
(\ref{Hamil inter dimless}) $\left\vert \psi(\xi)\right\rangle =D(-\gamma
)\left\vert \widetilde{\psi}\right\rangle $ in the following form%

\begin{equation}
\left\vert \psi(\xi)\right\rangle =%
{\displaystyle\sum\limits_{n=0}^{\infty}}
C_{n}(\widetilde{\gamma})\{\cos(\xi\sqrt{n+1})\left\vert e\right\rangle
\left\vert -\gamma;n\right\rangle -i\sin(\xi\sqrt{n+1})\left\vert
g\right\rangle \left\vert -\gamma;n+1\right\rangle \}\text{,}
\label{psi result}%
\end{equation}
where $\left\vert -\gamma;n\right\rangle =D(-\gamma)$\ $\left\vert
n\right\rangle $ is the displaced number state. Note the following. Formally
quantities $\mathcal{E}$ and $\alpha$ have different physical meaning.
Quantity $\mathcal{E}$ is used as a parameter of the driving field in the
Hamiltonian (\ref{Hamil full dimless}), while $\alpha$ is a factor of initial
condition for field's mode in the cavity. However dependence of probability
coefficients $C_{n}(\widetilde{\gamma})$ on the $\widetilde{\gamma
}=\mathcal{E}/2g+\alpha$ in (\ref{psi result}) shows a deep similarity of
these quantities for the resonant case $\delta=0$. The coherent field state
has minimum uncertainty, and resembles the classical field as closely as
quantum mechanics permits \cite{Glauber:1963a}. From the Eq.(\ref{psi result})
the density operator $\rho$\ can be written as follows

\begin{equation}
\rho=\left\vert \psi(\xi)\right\rangle \left\langle \psi(\xi)\right\vert
=\left\vert e\right\rangle \left\langle e\right\vert U_{ee}+\left\vert
e\right\rangle \left\langle g\right\vert U_{eg}+\left\vert g\right\rangle
\left\langle e\right\vert U_{ge}+\left\vert g\right\rangle \left\langle
g\right\vert U_{gg}\text{,} \label{Rho full}%
\end{equation}
where matrix elements $U_{ij}$ are given by%

\begin{align}
U_{ee}  &  =%
{\displaystyle\sum\limits_{n,m=0}^{\infty}}
C_{n}^{\ast}C_{m}\cos(\xi\sqrt{m+1})\cos(\xi\sqrt{n+1})\left\vert
-\gamma;m\right\rangle \left\langle -\gamma;n\right\vert \text{,}\label{Uab}\\
U_{eg}  &  =-i%
{\displaystyle\sum\limits_{n,m=0}^{\infty}}
C_{n}^{\ast}C_{m}\cos(\xi\sqrt{m+1})\sin(\xi\sqrt{n+1})\left\vert
-\gamma;m\right\rangle \left\langle -\gamma;n+1\right\vert \text{,}\nonumber\\
U_{eg}  &  =i%
{\displaystyle\sum\limits_{n,m=0}^{\infty}}
C_{n}^{\ast}C_{m}\cos(\xi\sqrt{m+1})\sin(\xi\sqrt{n+1})\left\vert
-\gamma;m+1\right\rangle \left\langle -\gamma;n\right\vert \text{,}\nonumber\\
U_{gg}  &  =%
{\displaystyle\sum\limits_{n,m=0}^{\infty}}
C_{n}^{\ast}C_{m}\sin(\xi\sqrt{m+1})\sin(\xi\sqrt{n+1})\left\vert
-\gamma;m+1\right\rangle \left\langle -\gamma;n+1\right\vert \text{,}\nonumber
\end{align}
and\ are still operators with respect to the field. To describe the evolution
of the atom (field) alone it is convenient to introduce the reduced density matrix%

\begin{equation}
\rho^{a(f)}=Tr_{f(a)}\{\rho\}\text{,} \label{pho_a}%
\end{equation}
where the trace is over the field (atom) states. We have used the subscript
$a,f$ \ to denote the atom (field). Unlike the state vector, the density
operator does not describe an individual system, but rather an ensemble of
identically prepared atoms, see e.g. Ref.\cite{Basdevant_Quantum:2002a}. A
condition for the ensemble to be in a pure state is that $Tr\{\left(
\rho^{a,f}\right)  ^{2}\}=1$. In this case a state-vector description of each
individual system of the ensemble is possible. On the other hand, for a
two-level system, a maximally mixed ensemble corresponds to $Tr\{\left(
\rho^{a,f}\right)  ^{2}\}=1/2$. Due to the identity%

\[
Tr_{f}\{\left\vert -\gamma;m\right\rangle \left\langle -\gamma;n\right\vert
\}=Tr_{a}\{D^{+}(\gamma)\left\vert m\right\rangle \left\langle n\right\vert
D(\gamma)\}=Tr_{a}\{\left\vert m\right\rangle \left\langle n\right\vert
D(\gamma)D^{+}(\gamma)\}=\delta_{mn}%
\]
one can write $\rho^{a}$ in the following form%

\begin{align}
\rho^{a}  &  =\{%
{\displaystyle\sum\limits_{n=1}^{\infty}}
\left\vert C_{n-1}\right\vert ^{2}\left(  \cos^{2}(\xi\sqrt{n})\left\vert
e\right\rangle \left\langle e\right\vert +\sin^{2}(\xi\sqrt{n})\left\vert
g\right\rangle \left\langle g\right\vert \right)  \}+\label{rho_a full}\\
&  +i\widetilde{\gamma}\{%
{\displaystyle\sum\limits_{n=1}^{\infty}}
\left\vert C_{n-1}\right\vert ^{2}\left(  \cos(\xi\sqrt{n+1})\sin(\xi\sqrt
{n})\left\vert e\right\rangle \left\langle g\right\vert -\sin(\xi\sqrt
{n+1})\cos(\xi\sqrt{n})\left\vert g\right\rangle \left\langle e\right\vert
\right)  \}\text{.}\nonumber
\end{align}
From Eq.(\ref{rho_a full}) one can see that off-diagonal matrix elements
$\rho_{eg}^{a}=\left\langle e\right\vert \rho^{a}\left\vert g\right\rangle $
and $\rho_{ge}^{a}=\left(  \rho_{eg}^{a}\right)  ^{+}$\ \ are of the first
order with respect to $\widetilde{\gamma}=\gamma+\alpha$, $\gamma
=\mathcal{E}/2g$ and therefore contain the information on the relative field's
phase even in a weak field limit. The mean photon number $\left\langle
n\right\rangle $ is given by%

\begin{equation}
\left\langle n\right\rangle =\left\langle a^{+}a\right\rangle _{f}%
=Tr\{a^{+}a\rho^{f}\}\text{.} \label{<n> init}%
\end{equation}
Taking into account the identity%

\[
Tr\{a^{+}a\left\vert -\gamma;m\right\rangle \left\langle -\gamma;n\right\vert
\}=Tr\{D^{+}Da^{+}D^{+}DaD^{+}(\gamma)\left\vert m\right\rangle \left\langle
n\right\vert D(\gamma)\}=\left\langle n\right\vert (a^{+}-\gamma^{\ast
})(a-\gamma)\left\vert m\right\rangle \text{,}%
\]
and after minor algebra one may write (\ref{<n> init} ) in the following form%

\begin{equation}
\left\langle n\right\rangle =\left\langle a^{+}a\right\rangle _{f}%
=A-B-B^{\ast}\text{,} \label{<n>}%
\end{equation}
where%

\begin{equation}
A=\left\vert \widetilde{\gamma}\right\vert ^{2}+\left\vert \gamma\right\vert
^{2}+%
{\displaystyle\sum\limits_{n=1}^{\infty}}
\left\vert C_{n-1}\right\vert ^{2}\sin^{2}(g\tau\sqrt{n})\text{,} \label{A}%
\end{equation}

\begin{equation}
B=\frac{1}{2}\gamma\widetilde{\gamma}^{\ast}%
{\displaystyle\sum\limits_{n=1}^{\infty}}
\left\vert C_{n-1}\right\vert ^{2}\frac{1}{\sqrt{n}}\{Q_{n}^{+}\cos\left[
Q_{n}^{-}g\tau\right]  -Q_{n}^{-}\cos\left[  Q_{n}^{+}g\tau\right]  \}\text{,}
\label{B}%
\end{equation}%
\[
\text{ }Q_{n}^{\pm}=\sqrt{n+1}\pm\sqrt{n}\text{, }C_{n}=\exp(-\left\vert
\widetilde{\gamma}\right\vert ^{2}/2)\frac{\widetilde{\gamma}^{n}}{\sqrt{n!}%
}\text{, }\widetilde{\gamma}=\gamma+\alpha=\frac{\mathcal{E}}{2g}%
+\alpha\text{.}%
\]
One can see, the quantum Rabi oscillations \ in (\ref{<n>}) appear as sum of
the sinusoidal terms at incommensurable frequencies $g\sqrt{n}$, $gQ_{n}^{\pm
}$, weighed by the probabilities $\left\vert C_{n}(\widetilde{\gamma
})\right\vert ^{2}$. For the vacuum case $\alpha=0$ equations (\ref{<n>}%
)-(\ref{B}) are in agreement with Ref.\cite{Chough:1996a}. One can see from
equations (\ref{A}), (\ref{B}) the following. Since $\left\vert C_{n}%
\right\vert ^{2}$ has maximum at $n\sim\left\vert \widetilde{\gamma
}\right\vert ^{2}$ ($n>1$) one may emphasize a desired frequency in the
spectrum the excited-state probability $P^{+}(\tau)=\rho_{ee}^{a}$
(\ref{rho_a full}) (the probability that the atom is in the excited state)
and\ $\left\langle n\right\rangle $\ (\ref{<n>}) by a corresponding choice of
complex quantity $\mathcal{E}$. Simple calculation yields the number of
corresponding Rabi frequency equals to $E[\widetilde{\gamma}^{2}-1]$, $E[x]$
is integer part of $\ x$.\ In other interesting case, the external field may
be chosen as%

\begin{equation}
\mathcal{E}=-2g\alpha\text{.} \label{E=-2ga}%
\end{equation}
In this case in equations (\ref{rho_a full}), (\ref{<n>}) $\widetilde{\gamma
}=0$, $B=0$, $C_{n}=\delta_{n1}$ , so $P^{+}$ and$\ \left\langle
n\right\rangle $ reduce to the following simple form%

\begin{equation}
P^{+}=\cos^{2}(g\tau)\text{ \ and }\left\langle n\right\rangle =\left\vert
\frac{\mathcal{E}}{2g}\right\vert ^{2}+\sin^{2}(g\tau)\text{.} \label{<n0>}%
\end{equation}
In this case the two-level atom, which is initially in the excited state,
undergoes the one-photon oscillations (radiating and absorption of a photon)
and only the vacuum Rabi frequency peak is present in the frequency spectrum.
Note the Rabi oscillations for standard JCM were directly observed in
Ref.\cite{Brune:1996b}.

The next experiment can be proposed on the basis of equations (\ref{E=-2ga}%
)-(\ref{<n0>}). One can vary both amplitude and phase of the external field
$\mathcal{E}$ until only single vacuum Rabi frequency spectrum is observed. In
this case condition $\alpha=-\mathcal{E}/2g$ must hold, which provides the
opportunity to measure parameters of the coherent state $\left\vert
\alpha\right\rangle $. In a compensative case (\ref{E=-2ga}) these fields have
equal amplitudes and frequency, but are shifted in phase by $\pi$. One can
interpret this as follows: in the resonant case ($\delta=0$) the total field
in the cavity may be written as a superposition of both fields in form
$\left(  \left\vert \alpha\right\rangle \pm\left\vert -\alpha\right\rangle
\right)  /\sqrt{2}$. It generates the Schr\"{o}dinger cat states since such
field is periodically entangled with a two-level atom while vacuum Rabi
oscillations occur. Method of generating such states have been explored by a
number of authors(see \cite{Yurke:1986a}, \cite{Brune:1996a},
\cite{Monroe:1996a}, \cite{Gerry:2002a} and references therein).

The above theory is valid in the case of very small dissipation (which
describes the interaction of the atom and field subsystems with the
environment) and zero detuning $\delta=0$. But in the experiments, the damping
of the cavity mode and the rate of spontaneous emission of the atom are not
negligibly small. Thus, a cavity damping must be included in a treatment of
driving JCM to compare it with experiments. An equation for density operator
$\rho(\tau)$ is required (master equation) because the loss of the coherence
due to the reservoir changes any system pure states to mixed states. With
cavity damping in effect we must solve master equation for the joint
atom-field density operator $\rho$ of the two-level atom coupled to the
electromagnetic field in cavity. Such equation is given by%

\begin{equation}
\frac{d\rho}{d\tau}=-i[H_{i},\rho]+%
\mathcal{L}%
_{1}\rho+%
\mathcal{L}%
_{2}\rho\text{,} \label{Neiman eq}%
\end{equation}
where $H_{i}$ is written in (\ref{full Hamil inter pict}). For simplicity we
only consider the case $\delta\neq0$ and $\Delta=0$. Master equation
(\ref{Neiman eq}) is more difficult to solve and numerical methods usually
must be used. At interesting frequencies range the dissipation is written in
Eq.(\ref{Neiman eq}) both as mirror losses in the cavity that defines the mode
of the electromagnetic field $%
\mathcal{L}%
_{1}\rho$, and as spontaneous emission from the atom $%
\mathcal{L}%
_{2}\rho$. At Born-Markov approximation and zero temperature, these parts are
written in equation (\ref{Neiman eq}) as the following. One term is as follows
$%
\mathcal{L}%
_{1}\rho=\gamma_{1}\left(  2a\rho a^{+}-a^{+}a\rho-\rho a^{+}a\right)  $,
where $\gamma_{1}$ is the rate of single-photon losses. Another term $%
\mathcal{L}%
_{2}\rho=\left(  \gamma_{2}/2\right)  \left(  2\sigma\rho\sigma^{+}-\sigma
^{+}\sigma\rho-\rho\sigma^{+}\sigma\right)  $ takes into account the
spontaneous emission from the atom out of the sides of the
cavity\cite{Scully_Quantum:1996a}. In this case the atom is damped by
spontaneous emission with damping rate $\gamma_{2}$ to modes other than the
privileged cavity mode with frequency $\omega_{a}$. We have solved the
equation (\ref{Neiman eq}) numerically using a truncated number states
$\left\vert n\right\rangle $ and atom states $\left\vert e\right\rangle $,
$\left\vert g\right\rangle $ basis. The algorithms for integration of such a
system numerically can be found, e.g. in Ref.\cite{Press_Numerical:2002a}. In
general $0\leq n\leq\infty$, but for numerical calculations we have used
$0\leq n\leq M$. A finite base of the number states $M$ was kept large enough
so that the highest energy state is never populated significantly. Since we
assume above that the field is initially in the coherent state $\left\vert
\alpha\right\rangle $, and the atom is in the upper state $\left\vert
e\right\rangle $, then initially $\rho=\left\vert \alpha\right\rangle
\left\langle \alpha\right\vert \otimes\left\vert e\right\rangle \left\langle
e\right\vert $ must hold. Using numerically obtained joint density matrix
$\rho$ \ we have calculated the next quantities. Tracing out the matrix $\rho$
over the field (atom) states we have calculated the reduced atom (field)
density matrix $\rho^{a,f}=Tr_{f,a}\{\rho\}$ , where $Tr_{a,f}\{\rho\}$ are
the partial traces over the atom or field states accordingly. The latter
allows us to study the dynamics of the excited-state probability $P^{+}%
(\tau)=\rho_{ee}^{a}$ , the mean photon number $\left\langle n(\tau
)\right\rangle $ and the entropy $S^{a}(\tau)=-Tr\{\rho^{a}\ln\rho^{a}\}$.
Also the Fourier spectrum of $P^{+}(\tau)$ is explored. The convergence of the
equations is tested and the dynamics of the system is studied for different
values of the external field relative to the atom-field mode coupling. The
results of the numerical solution of master equation (\ref{Neiman eq}) are
shown in Figs.\ref{pic_fig1}-\ref{pic_fig4}.

\section{Numerical results}

\begin{figure}
[ptb]
\begin{center}
\includegraphics[
height=4.2211in,
width=5.2667in
]%
{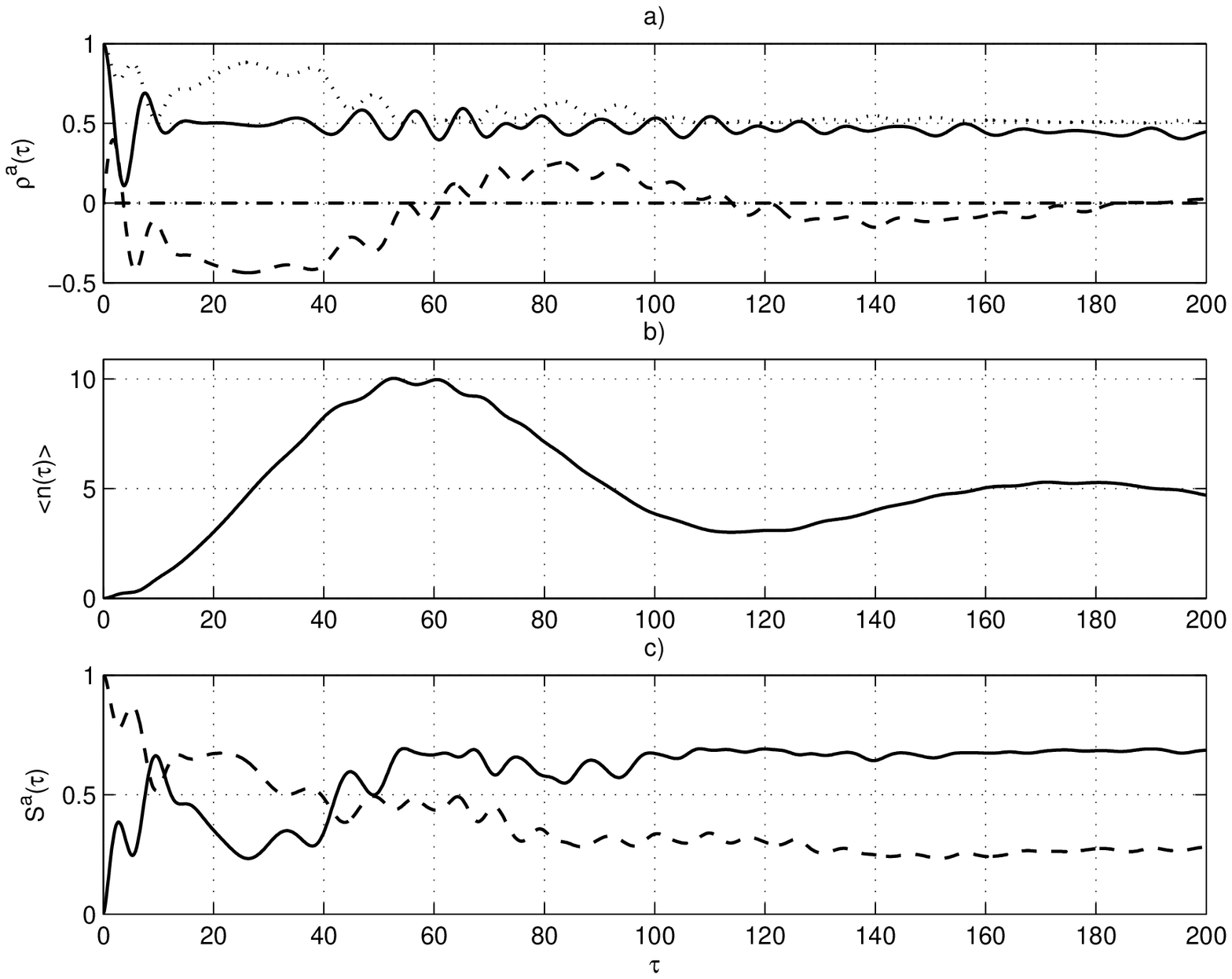}%
\caption{Quantum dynamics in driving JCM as function of the interaction time
$\tau$\ for initial vacuum case $\left\vert 0\right\rangle $, $\delta
=\Delta=0$, $\mathcal{E}=0.7$, $g=0.2$, $\gamma_{1}=5\cdot10^{-3}$,
$\gamma_{2}=10^{-3}$. (a) Probability of the excited level $\left\vert
e\right\rangle $\ occupations $P^{+}$ (solid line), $\operatorname{Im}%
(\rho_{ge}^{a})$ (dashed line), $Tr\{(\rho^{a})^{2}\}$ (dotted line). (b) Mean
photon number $\left\langle n\right\rangle $. (c) Entropy $S^{a}$ (solid
line), $Tr\{(\rho^{f})^{2}\}$ (dashed line). See details in text.}%
\label{pic_fig1}%
\end{center}
\end{figure}

We briefly consider the dynamics of a vacuum $\alpha=0$ of driving JCM in
order to understand some general features first. Fig.\ref{pic_fig1} shows the
behavior of the atom quantities and mean photon number for resonant case
($\delta=0$) for vacuum initial state $\left\vert 0\right\rangle $ at losses
case, obtained as a result of the numerical solution of the master equation
(\ref{Neiman eq}). This solution is close to Eq.(\ref{rho_a full}), but with a
damping due to both mirror losses in the cavity and spontaneous emission from
the atom. We use the following parameters $\mathcal{E=}0.7$, $g=0.2$,
$\gamma_{1}/\gamma_{2}=5$ and $\gamma_{1}=5\cdot10^{-3}$. To make clearer the
details of time evolution we use here a time interval $\tau=200$ , which is
large in comparison with the characteristic time scale of
revival\cite{Narozhny:1981a} $\tau>2\tau_{R}$, where $\tau_{R}=2\gamma\pi/g$
$=55$, $\gamma=\mathcal{E}/2g=1.75$. One can see, that dynamics of the
probability of excited level occupations $P^{+}(\tau)$ has form of a damped
sequence of collapses and revivals. Dynamical collapses and revivals are
specific features of a unitary evolution. They are strongly affected by
decoherency, which have the time constant set by the cavity field energy
damping time $\sim1/(\gamma_{1}+\gamma_{2})$. From Fig.\ref{pic_fig1}(a) one
can see that both collapses and revivals are progressively less pronounced due
to dissipation $\gamma_{1,2}\neq0$. The dissipation also reduces the
magnitudes of non-diagonal elements $\rho_{ge}^{a}$ (in this case $\rho
_{eg}^{a}=(\rho_{ge}^{a})^{\ast}$ are purely imaginary quantities). One can
see, that quantity $\operatorname{Im}(\rho_{ge}^{a})$ over time approaches to
zero due to losses, which causes the field phase information to wash out and
progresses decoherence in the coupled system.

The dynamics $Tr\{(\rho^{a})^{2}\}$ in Fig.\ref{pic_fig1}(a)\ \ shows that in
the collapse area quantity $Tr\{(\rho^{a})^{2}\}$ has a maximum close to $1$
at $\tau_{0}=\tau_{R}/2=27.5$. This point corresponds to the atomic attractor
state which is completely independent on the initial atomic
state\cite{JulioGea-Banacloche:1990a}. At this point the compound system is in
the disentangled state when in lossless system $S^{a,f}\rightarrow0$ and
$Tr\{(\rho^{a,f})^{2}\}\rightarrow1$. In lossy system such limit values are
fulfilled only approximately. One can see from Fig.\ref{pic_fig1}(b) that the
mean photons number $\left\langle n\right\rangle $\ increases from the initial
zero value; however over time $\left\langle n\right\rangle $\ assumes the
steady-state value justified by the amplitude of external driving field
$\mathcal{E}$. Fig.\ref{pic_fig1}(c) shows the dynamics of the entropy of atom
subsystem $S^{a}(\tau)=-Tr\{\rho^{a}\ln\rho^{a}\}$. In the area of the first
collapse $S^{a}$ has oscillating behavior. However these oscillations are
smoothed away over time and $S^{a}(\tau)$ approaches to the steady-state value
$ln(2)=0.69$, which corresponds to maximally entangled state of two-level atom
and field mode. At a lossless case the exact equalities $Tr\{(\rho^{a}%
)^{2}\}=Tr\{(\rho^{f})^{2}\}$ and $S^{a}=S^{f}$ in standard JCM hold
\cite{Phoenix:1991a}. Our simulations confirm this fact with very good accuracy.

\begin{figure}
[ptb]
\begin{center}
\includegraphics[
height=4.2214in,
width=5.2663in
]%
{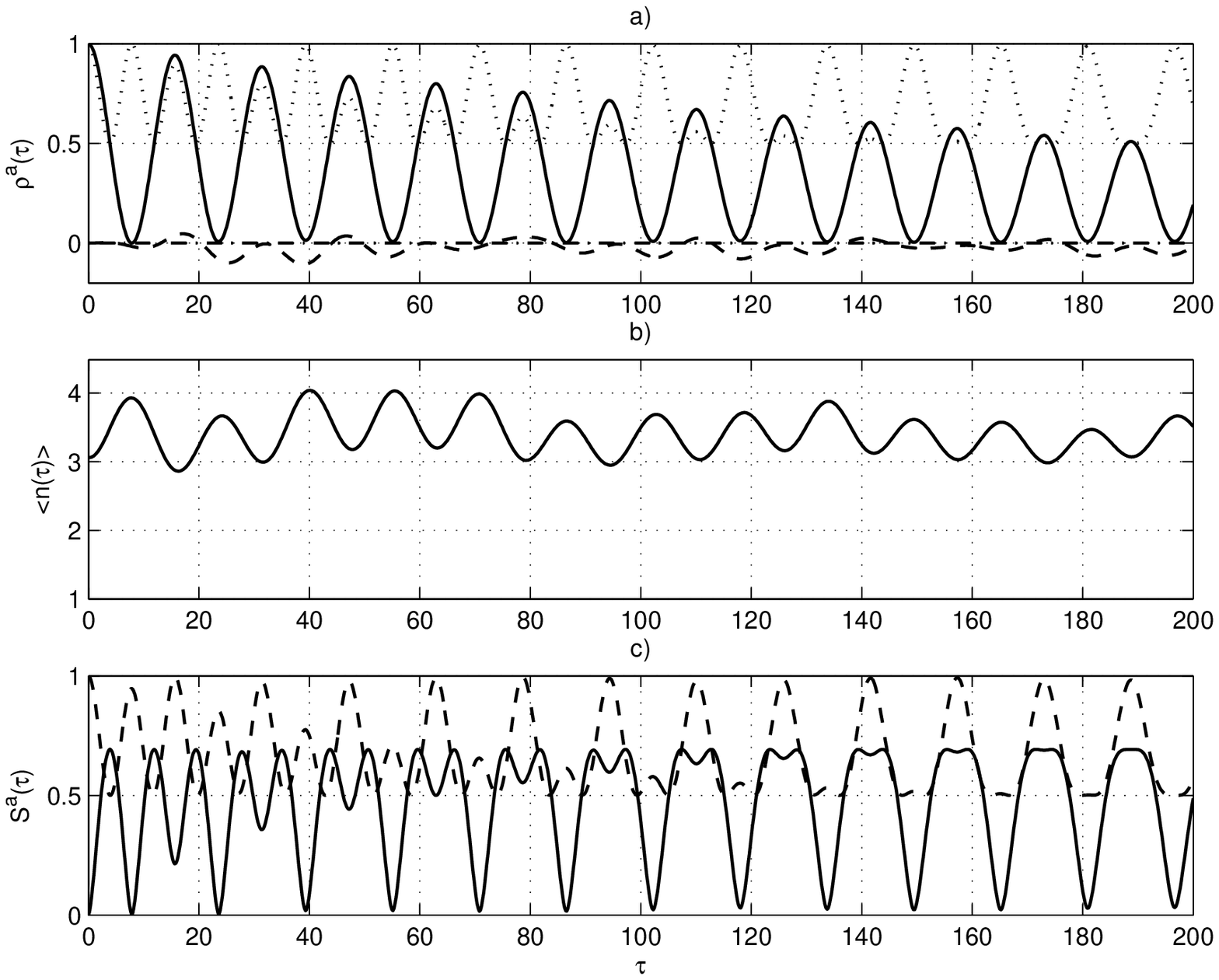}%
\caption{Quantum dynamics in driving JCM as function of the interaction time
$\tau$\ for initial coherent case $\left\vert \alpha\right\rangle $,
$\alpha=-1.75$ (compensating case (\ref{E=-2ga})), $\delta=\Delta=0$,
$\mathcal{E}=0.7$, $g=0.2$, $\gamma_{1}=5\cdot10^{-3}$, $\gamma_{2}=10^{-3}$.
(a) Probability of the excited level $\left\vert e\right\rangle $\ occupations
$P^{+}$ (solid line), $\operatorname{Im}(\rho_{ge}^{a})$ (dashed line),
$\operatorname{Re}(\rho_{ge}^{a})$ (dash-dot line), $Tr\{(\rho^{a})^{2}\}$
(dotted line). (b) Mean photon number $\left\langle n\right\rangle $. (c)
Entropy $S^{a}$ (solid line), $Tr\{(\rho^{f})^{2}\}$ (dashed line). See
details in text.}%
\label{pic_fig2}%
\end{center}
\end{figure}

Next, we study details of the driving and initial coherent fields interaction.
Fig.\ref{pic_fig2}(a) shows dynamics of the atom subsystem for the
compensative case ($\widetilde{\gamma}=\gamma+\alpha=\mathcal{E}/2g+\alpha=0$)
( see Eq. (\ref{E=-2ga}) for the exact resonance case ($\delta=0$)), but
taking into account dissipation $\gamma_{1,2}\neq0$. Comparison of
Fig.\ref{pic_fig1}(a) and Fig.\ref{pic_fig2}(a) shows that the dynamics for
this case is essentially different from the that in the initial vacuum case
$\alpha=0$ and has the form of damped vacuum Rabi oscillation in standard JCM.

Dynamics of the mean photons number $\left\langle n\right\rangle $ is of
interest and is shown in Fig.\ref{pic_fig2}(b). The solid line in
Fig.\ref{pic_fig2}(b) shows results $\left\langle n\right\rangle $ from a
numerical simulation of the atom-field master equation (\ref{Neiman eq}) for
the compensative case, taking into account losses in the system. For a short
period of time $\tau<10$ this simulation is in very good agreement with the
exact formula (\ref{<n0>}). However a discrepancy arises over time due to
losses not included in (\ref{<n0>}). One can see that despite of the mutual
compensation of the initial coherent field and driving field in Eq.(\ref{<n0>}%
), the mean number of photons oscillates in vicinity $\sim\left\vert
\alpha\right\vert ^{2}$, that is justified by the field $\left\vert
\mathcal{E}\right\vert $. One can see from Fig.\ref{pic_fig2}(c), that for a
short time the dynamics of entropy is similar to that in the vacuum case.
However over longer time intervals the impact of dissipation becomes
essential. Despite dissipation, at the time instances $\tau_{k}=\pi(2k+1)/2g$
, $k=0,1...$ the entropy $S^{a}(\tau_{k})$ is very close to zero that implies
the transition of the coupled atom-field system to the uncoupled pure state.
At the moments $\tau_{k}=\pi k/g$ the entropy has a value close to $\ln2$,
which corresponds to the maximum entanglement of the two-level atom and the
field. Since the mean photon number does not vanish (see Fig.\ref{pic_fig2}%
(b)) this dynamics is stable asymptotically.

\begin{figure}
[ptb]
\begin{center}
\includegraphics[
height=4.2211in,
width=5.1923in
]%
{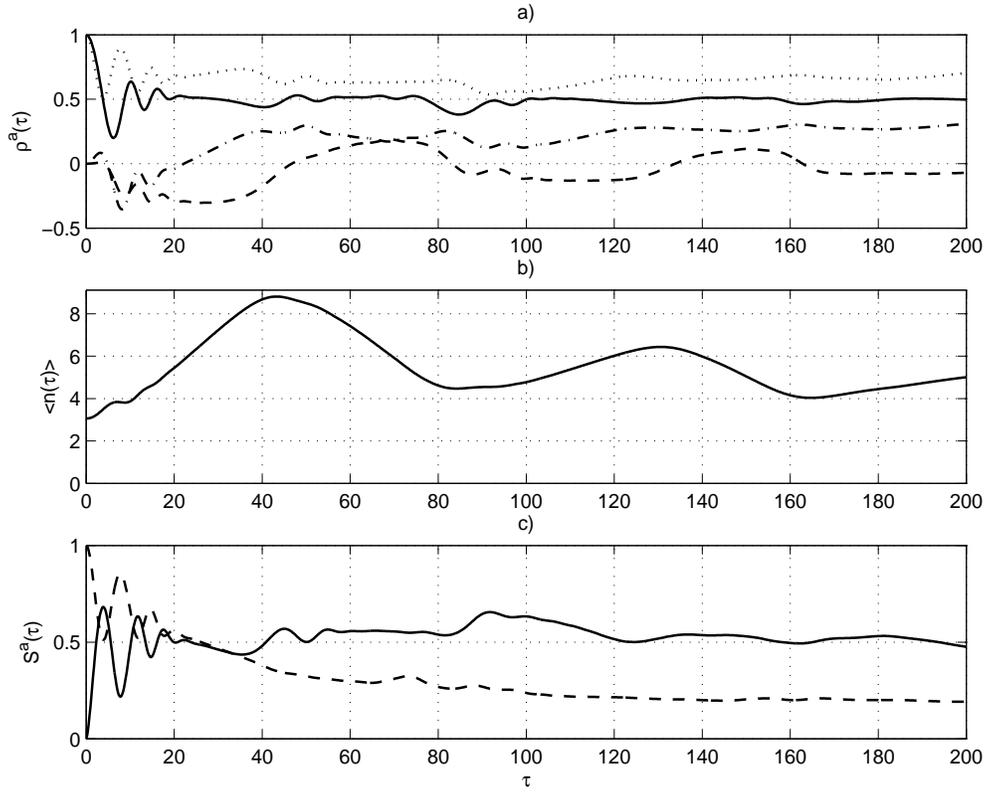}%
\caption{The same as in Fig.\ref{pic_fig2} but for detuning $\delta=0.1$.}%
\label{pic_fig3}%
\end{center}
\end{figure}

Fig.\ref{pic_fig3} shows dynamics of the coupled subsystem in the compensative
case ($\widetilde{\gamma}=0$) when both dissipation $\gamma_{1},\gamma_{2}%
\neq0$ and detuning $\delta\neq0$ are non-zero. Notice in a non-resonant case
($\delta\neq0$) the time dependence in the Hamiltonian
(\ref{full Hamil inter pict}) is not removed, and therefore the displacement
operators transformation already does not allow to reduce the driving JCM to
standard JCM. Compared to the case of the exact resonance (Fig.\ref{pic_fig2})
the sequential collapses and revivals now practically disappear over long
periods of time; in Fig.\ref{pic_fig3}(a) the first collapse is seen only.
Nevertheless some maxima of $Tr\{(\rho^{a})^{2}\}$ remain resolvable. One can
see from Fig.\ref{pic_fig3}(c) that during short periods of time $\tau<7$\ the
behavior $S^{a}$ is similar to a that in the vacuum case (Fig.\ref{pic_fig2}%
(c)). However over longer periods of time the influence of decoherence becomes
essential. The entropy quickly approaches its asymptotic value.

\begin{figure}
[ptb]
\begin{center}
\includegraphics[
height=4.2523in,
width=5.1629in
]%
{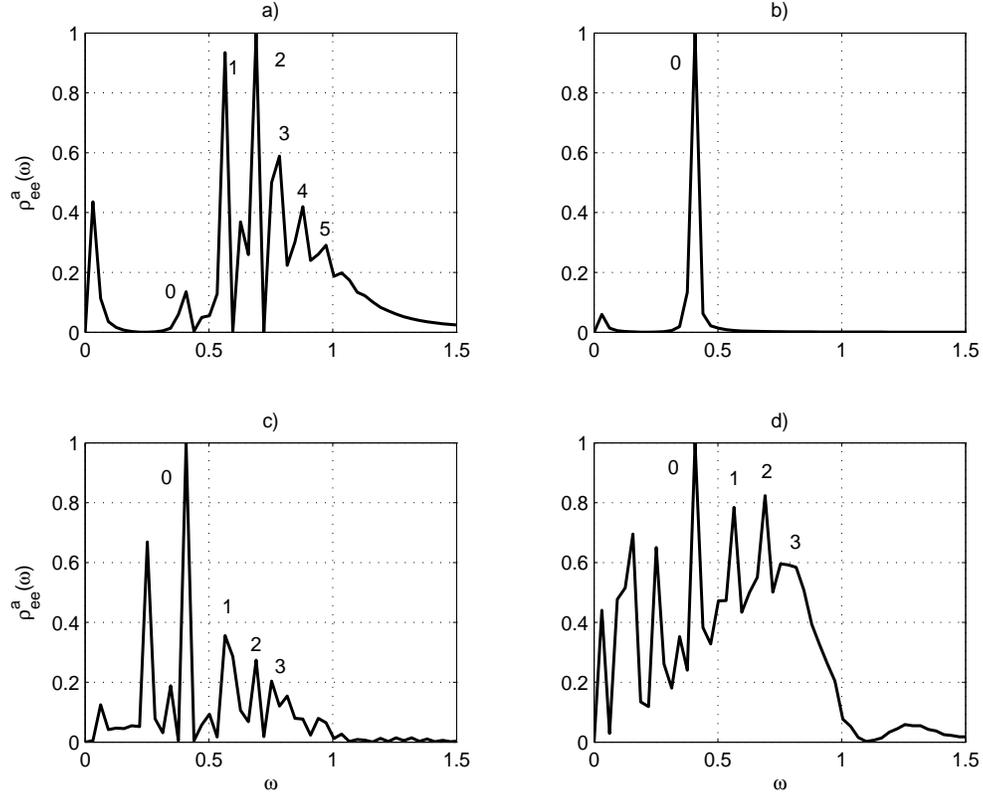}%
\caption{Fourier transforms of the probability $P^{+}(\tau)$ revealing the
discrete Rabi frequencies, occurring at the successive square roots of the
integers $\sqrt{n+1}$, numbers $n=0,1,2...$ are inserted in vicinity of peaks.
Spectra correspond to (a) Fig.\ref{pic_fig1}(a); (b) Fig.\ref{pic_fig2}(a);
(c) Fig.\ref{pic_fig3}(a), but with $\gamma_{1}=5\cdot10^{-5}$, $\gamma
_{2}=10^{-5}$; (d) Fig.\ref{pic_fig3}(a).}%
\label{pic_fig4}%
\end{center}
\end{figure}

Fig.\ref{pic_fig4} shows the Fourier spectra for cases of driving JCM
considered above. This spectrum, obtained by Fourier transform of numerically
calculated $P^{+}(\tau)$, exhibits well separated discrete frequency
components, which are scaled as square roots of the successive integers. The
spectrum in Fig.\ref{pic_fig4}(a) corresponds to the time dynamics presented
in Fig.\ref{pic_fig1}(a). One can see, that $P^{+}$ spectrum at the initial
vacuum state $\left\vert 0\right\rangle $ for driving JCM case\ is similar to
that in a standard JCM (without driving field $\mathcal{E}=0$) at initial
coherent field $\left\vert \alpha\right\rangle $. Note that for standard JCM
such peaks were observed experimentally in Ref.\cite{Brune:1996b}. The
spectrum in Fig.\ref{pic_fig4} (a) is rather similar to the spectrum shown in
Fig.2(d) of experiment\cite{Brune:1996b} for $\gamma=1.77$. We can conclude
that the vacuum case of driving JCM is similar to a standard JCM case with a
coherent initial field, at least for the exact resonance. This conclusion
reiterates the fact that coherent field state is as close to the classical
field as quantum mechanics permits\cite{Glauber:1963a}. Note that
Fig.\ref{pic_fig4}(a) shows that peak number $2$ is the highest one. This also
follows from the above mentioned. In this case $\widetilde{\gamma}=1.75$
therefore the calculated number of highest peak is $E[\widetilde{\gamma}%
^{2}-1]=2$.

Figs.\ref{pic_fig4}(b)-\ref{pic_fig4}(d) represent compensative case
(\ref{E=-2ga}) for the loss and detuning case. Fig.\ref{pic_fig4}(b) shows the
spectrum for case $\delta=0$, however for not small losses, which are. One can
see only the peak, which corresponds to $n=0$ is present in the Rabi spectrum.
Such spectrum in Fig.\ref{pic_fig4} (b) corresponds to the measured spectrum
shown in Fig.2 (a) in Ref.\cite{Brune:1996b} \ for nearly vacuum case (no
injected fields). Thus, the conclusion about the possibility subtracting of
the coherent and classical fields remains valid in the loss case also. In
Fig.\ref{pic_fig4}(c) influence of detuning is shown for smaller losses. One
can see, that even for $\delta=0.1$ which is not very small the spectrum of
several first Rabi frequencies is well recognizable. There are a few peaks in
Fig.\ref{pic_fig4}(c), which correspond to detuning $\delta\neq0$. Such peaks
are present even at a larger dissipation rate (the dissipation rate in
Fig.\ref{pic_fig4}(d) are increased by two orders, with respect to
Fig.\ref{pic_fig4}(c). Nevertheless the peak of vacuum oscillations remains dominant.

In general one can readily derive, that the quantity $\gamma$ can be rewritten
(with the previously used notations) as\ $\gamma=\mathcal{E}/2g=E_{e}/E_{vac}%
$, where $E_{vac}=\left(  \hbar\omega_{a}/2\varepsilon_{0}V\right)  ^{1/2}%
$\ is the field per photon. Then from equality (\ref{E=-2ga}) $\widetilde
{\gamma}=\gamma+\alpha=0$ one can obtain $E_{e}=E_{vac}\overline{n}^{1/2}%
\exp(i(v+\pi))$. The latter equation states that for the compensated case
(\ref{E=-2ga}) the driving field $E_{e}$\ should be in a coherent state, but
shifted by phase $\pi$ concerning to initial state $\alpha$. In general,
condition (\ref{E=-2ga}) does not hold exactly due to quantum fluctuations.
However, since the ratio of fluctuation to the mean number (fractional
uncertainty in the photon number, see e.g. Ch.3 in
Ref.\cite{ChristopherGerry_Introductory:2004a}) is $\Delta n/\overline
{n}=\overline{n}^{-1/2}$ for a Poisson process, the larger values of
$\overline{n}$ become, the better is the accuracy of condition (\ref{E=-2ga}%
).\bigskip

\section{Conclusion}

Studied the field-atom interactions in the driving JCM shows that a variation
of the amplitude or/and the phase of the driving field enables one to
manipulate the dynamics and the spectrum of quantum Rabi oscillations. There
are two distinct regimes. In case of summation of the driving field and
initial coherent field one can underscoring a selected frequency in the Rabi
frequency spectrum. The subtraction provides a possibility of compensation of
both fields. For the case of the exact compensation, the frequency spectrum of
a two-level atom becomes similar to that of vacuum oscillations in standard
JCM. To use these processes for quantum information technology, the
decoherence time must be greater than the time scale of the atom-field
interaction. In this case, these processes may have various applications
related to monitoring of the entanglement of two-level atom with quantized
field and may be used to develop quantum information technology devices.

\section{Acknowledgements}

The author is grateful to Alexander (Sasha) Draganov (ITT Industries -
Advanced Engineering \& Science Division) who made several helpful comments.

\bigskip

\input{BurlPLA.bbl}

\newpage%

\newpage%

\newpage

\bigskip%

\newpage

\bigskip%

\newpage

\bigskip
\end{document}

%% file: BurlPLA.bbl
\newif\ifabfull\abfulltrue

%% file: Burlak.bbl
\begin{thebibliography}{28}
\expandafter\ifx\csname natexlab\endcsname\relax\def\natexlab#1{#1}\fi
\expandafter\ifx\csname bibnamefont\endcsname\relax
  \def\bibnamefont#1{#1}\fi
\expandafter\ifx\csname bibfnamefont\endcsname\relax
  \def\bibfnamefont#1{#1}\fi
\expandafter\ifx\csname citenamefont\endcsname\relax
  \def\citenamefont#1{#1}\fi
\expandafter\ifx\csname url\endcsname\relax
  \def\url#1{\texttt{#1}}\fi
\expandafter\ifx\csname urlprefix\endcsname\relax\def\urlprefix{URL }\fi
\providecommand{\bibinfo}[2]{#2}
\providecommand{\eprint}[2][]{\url{#2}}

\bibitem[{\citenamefont{Blinov et~al.}(2004)\citenamefont{Blinov, Moehring,
  Duan, and Monroe}}]{Blinov:2004a}
\bibinfo{author}{\bibfnamefont{B.~B.} \bibnamefont{Blinov}},
  \bibinfo{author}{\bibfnamefont{D.~L.} \bibnamefont{Moehring}},
  \bibinfo{author}{\bibfnamefont{L.-M.} \bibnamefont{Duan}}, \bibnamefont{and}
  \bibinfo{author}{\bibfnamefont{C.}~\bibnamefont{Monroe}},
  \bibinfo{journal}{Nature.} \textbf{\bibinfo{volume}{428}},
  \bibinfo{pages}{153 } (\bibinfo{year}{2004}).

\bibitem[{\citenamefont{Zurek}(2003)}]{WojciechHubertZurek:2003a}
\bibinfo{author}{\bibfnamefont{W.~H.} \bibnamefont{Zurek}},
  \bibinfo{journal}{Rev. Mod. Phys.} \textbf{\bibinfo{volume}{75}},
  \bibinfo{pages}{715} (\bibinfo{year}{2003}).

\bibitem[{\citenamefont{Raimond et~al.}(2001)\citenamefont{Raimond, Brune, and
  Haroche}}]{Raimond:2001a}
\bibinfo{author}{\bibfnamefont{J.~M.} \bibnamefont{Raimond}},
  \bibinfo{author}{\bibfnamefont{M.}~\bibnamefont{Brune}}, \bibnamefont{and}
  \bibinfo{author}{\bibfnamefont{S.}~\bibnamefont{Haroche}},
  \bibinfo{journal}{Rev. Mod. Phys.} \textbf{\bibinfo{volume}{73}},
  \bibinfo{pages}{565} (\bibinfo{year}{2001}).

\bibitem[{\citenamefont{Brune et~al.}(1996{\natexlab{a}})\citenamefont{Brune,
  Hagley, Dreyer, Maitre, Maali, Wunderlich, Raimond, and
  Haroche}}]{Brune:1996a}
\bibinfo{author}{\bibfnamefont{M.}~\bibnamefont{Brune}},
  \bibinfo{author}{\bibfnamefont{E.}~\bibnamefont{Hagley}},
  \bibinfo{author}{\bibfnamefont{J.}~\bibnamefont{Dreyer}},
  \bibinfo{author}{\bibfnamefont{X.}~\bibnamefont{Maitre}},
  \bibinfo{author}{\bibfnamefont{A.}~\bibnamefont{Maali}},
  \bibinfo{author}{\bibfnamefont{C.}~\bibnamefont{Wunderlich}},
  \bibinfo{author}{\bibfnamefont{J.~M.} \bibnamefont{Raimond}},
  \bibnamefont{and} \bibinfo{author}{\bibfnamefont{S.}~\bibnamefont{Haroche}},
  \bibinfo{journal}{Phys. Rev. Lett.} \textbf{\bibinfo{volume}{77}},
  \bibinfo{pages}{4887} (\bibinfo{year}{1996}{\natexlab{a}}).

\bibitem[{\citenamefont{Jaynes and Cummings}(1963)}]{Jaynes:1963a}
\bibinfo{author}{\bibfnamefont{E.}~\bibnamefont{Jaynes}} \bibnamefont{and}
  \bibinfo{author}{\bibfnamefont{F.}~\bibnamefont{Cummings}},
  \bibinfo{journal}{Proc. IEEE.} \textbf{\bibinfo{volume}{51}},
  \bibinfo{pages}{89} (\bibinfo{year}{1963}).

\bibitem[{\citenamefont{Solano et~al.}(2003)\citenamefont{Solano, Agarwal, and
  Walther}}]{Solano:2003a}
\bibinfo{author}{\bibfnamefont{E.}~\bibnamefont{Solano}},
  \bibinfo{author}{\bibfnamefont{G.~S.} \bibnamefont{Agarwal}},
  \bibnamefont{and} \bibinfo{author}{\bibfnamefont{H.}~\bibnamefont{Walther}},
  \bibinfo{journal}{Phys. Rev. Lett.} \textbf{\bibinfo{volume}{90}},
  \bibinfo{pages}{027903} (\bibinfo{year}{2003}).

\bibitem[{\citenamefont{Zheng}(2002)}]{Shi-BiaoZheng:2002a}
\bibinfo{author}{\bibfnamefont{S.-B.} \bibnamefont{Zheng}},
  \bibinfo{journal}{Phys. Rev. A.} \textbf{\bibinfo{volume}{66}},
  \bibinfo{pages}{060303 (4 pages)} (\bibinfo{year}{2002}).

\bibitem[{\citenamefont{Zheng}(2003)}]{Shi-BiaoZheng:2003a}
\bibinfo{author}{\bibfnamefont{S.-B.} \bibnamefont{Zheng}},
  \bibinfo{journal}{Phys. Rev. A.} \textbf{\bibinfo{volume}{68}},
  \bibinfo{pages}{035801 (4 pages)} (\bibinfo{year}{2003}).

\bibitem[{\citenamefont{Lougovski et~al.}(2004)\citenamefont{Lougovski,
  Casagrande, Lulli, Englert, Solano, and Walther}}]{Lougovski:2004a}
\bibinfo{author}{\bibfnamefont{P.}~\bibnamefont{Lougovski}},
  \bibinfo{author}{\bibfnamefont{F.}~\bibnamefont{Casagrande}},
  \bibinfo{author}{\bibfnamefont{A.}~\bibnamefont{Lulli}},
  \bibinfo{author}{\bibfnamefont{B.-G.} \bibnamefont{Englert}},
  \bibinfo{author}{\bibfnamefont{E.}~\bibnamefont{Solano}}, \bibnamefont{and}
  \bibinfo{author}{\bibfnamefont{H.}~\bibnamefont{Walther}},
  \bibinfo{journal}{Phys. Rev. A.} \textbf{\bibinfo{volume}{69}},
  \bibinfo{pages}{023812 (9 pages)} (\bibinfo{year}{2004}).

\bibitem[{\citenamefont{Vahala}(2003)}]{Vahala:2003a}
\bibinfo{author}{\bibfnamefont{K.~J.} \bibnamefont{Vahala}},
  \bibinfo{journal}{Nature.} \textbf{\bibinfo{volume}{424}},
  \bibinfo{pages}{839} (\bibinfo{year}{2003}).

\bibitem[{\citenamefont{Artemyev et~al.}(2001)\citenamefont{Artemyev, Woggon,
  and Wannemacher}}]{Artemyev:2001a}
\bibinfo{author}{\bibfnamefont{M.~V.} \bibnamefont{Artemyev}},
  \bibinfo{author}{\bibfnamefont{U.}~\bibnamefont{Woggon}}, \bibnamefont{and}
  \bibinfo{author}{\bibfnamefont{R.}~\bibnamefont{Wannemacher}},
  \bibinfo{journal}{Appl. Phys. Lett.} \textbf{\bibinfo{volume}{78}},
  \bibinfo{pages}{1032} (\bibinfo{year}{2001}).

\bibitem[{\citenamefont{G.Burlak et~al.}(2003)\citenamefont{G.Burlak, Marquez,
  and O.Starostenko}}]{Burlak:2003a}
\bibinfo{author}{\bibnamefont{G.Burlak}}, \bibinfo{author}{\bibfnamefont{P.~A.}
  \bibnamefont{Marquez}}, \bibnamefont{and}
  \bibinfo{author}{\bibnamefont{O.Starostenko}}, \bibinfo{journal}{Phys. Lett.
  A.} \textbf{\bibinfo{volume}{309}}, \bibinfo{pages}{146}
  (\bibinfo{year}{2003}).

\bibitem[{\citenamefont{Alsing et~al.}(1992)\citenamefont{Alsing, Guo, and
  Carmichael}}]{Alsing:1992a}
\bibinfo{author}{\bibfnamefont{P.}~\bibnamefont{Alsing}},
  \bibinfo{author}{\bibfnamefont{D.-S.} \bibnamefont{Guo}}, \bibnamefont{and}
  \bibinfo{author}{\bibfnamefont{H.~J.} \bibnamefont{Carmichael}},
  \bibinfo{journal}{Phys. Rev. A.} \textbf{\bibinfo{volume}{45}},
  \bibinfo{pages}{5135} (\bibinfo{year}{1992}).

\bibitem[{\citenamefont{Jyotsna and Agarwal}(1993)}]{Jyotsna:1993a}
\bibinfo{author}{\bibfnamefont{I.~V.} \bibnamefont{Jyotsna}} \bibnamefont{and}
  \bibinfo{author}{\bibfnamefont{G.~S.} \bibnamefont{Agarwal}},
  \bibinfo{journal}{Opt. Commun.} \textbf{\bibinfo{volume}{99}},
  \bibinfo{pages}{344} (\bibinfo{year}{1993}).

\bibitem[{\citenamefont{Dutra et~al.}(1993)\citenamefont{Dutra, Knight, and
  Moya-Cessa}}]{Dutra:1993a}
\bibinfo{author}{\bibfnamefont{S.~M.} \bibnamefont{Dutra}},
  \bibinfo{author}{\bibfnamefont{P.~L.} \bibnamefont{Knight}},
  \bibnamefont{and}
  \bibinfo{author}{\bibfnamefont{H.}~\bibnamefont{Moya-Cessa}},
  \bibinfo{journal}{Phys. Rev. A.} \textbf{\bibinfo{volume}{48}},
  \bibinfo{pages}{3168} (\bibinfo{year}{1993}).

\bibitem[{\citenamefont{Chough and Carmichael}(1996)}]{Chough:1996a}
\bibinfo{author}{\bibfnamefont{Y.~T.} \bibnamefont{Chough}} \bibnamefont{and}
  \bibinfo{author}{\bibfnamefont{H.~J.} \bibnamefont{Carmichael}},
  \bibinfo{journal}{Phys. Rev. A.} \textbf{\bibinfo{volume}{54}},
  \bibinfo{pages}{1709} (\bibinfo{year}{1996}).

\bibitem[{\citenamefont{Gerry}(2002)}]{Gerry:2002a}
\bibinfo{author}{\bibfnamefont{C.~C.} \bibnamefont{Gerry}},
  \bibinfo{journal}{Phys. Rev.} \textbf{\bibinfo{volume}{A 65}},
  \bibinfo{pages}{063801 (6 pages)} (\bibinfo{year}{2002}).

\bibitem[{\citenamefont{Glauber}(1963)}]{Glauber:1963a}
\bibinfo{author}{\bibfnamefont{R.}~\bibnamefont{Glauber}},
  \bibinfo{journal}{Phys. Rev.} \textbf{\bibinfo{volume}{131}},
  \bibinfo{pages}{2766} (\bibinfo{year}{1963}).

\bibitem[{\citenamefont{J.-L.Basdevant and
  J.Dalibard}(2002)}]{Basdevant_Quantum:2002a}
\bibinfo{author}{\bibnamefont{J.-L.Basdevant}} \bibnamefont{and}
  \bibinfo{author}{\bibnamefont{J.Dalibard}}, \emph{\bibinfo{title}{{Quantum
  mechanics}}} (\bibinfo{publisher}{Springer}, \bibinfo{year}{2002}).

\bibitem[{\citenamefont{Brune et~al.}(1996{\natexlab{b}})\citenamefont{Brune,
  Schmidt-Kaler, Maali, Dreyer, Hagley, Raimond, and Haroche}}]{Brune:1996b}
\bibinfo{author}{\bibfnamefont{M.}~\bibnamefont{Brune}},
  \bibinfo{author}{\bibfnamefont{F.}~\bibnamefont{Schmidt-Kaler}},
  \bibinfo{author}{\bibfnamefont{A.}~\bibnamefont{Maali}},
  \bibinfo{author}{\bibfnamefont{J.}~\bibnamefont{Dreyer}},
  \bibinfo{author}{\bibfnamefont{E.}~\bibnamefont{Hagley}},
  \bibinfo{author}{\bibfnamefont{J.~M.} \bibnamefont{Raimond}},
  \bibnamefont{and} \bibinfo{author}{\bibfnamefont{S.}~\bibnamefont{Haroche}},
  \bibinfo{journal}{Phys. Rev. Lett.} \textbf{\bibinfo{volume}{76}},
  \bibinfo{pages}{1800–1803} (\bibinfo{year}{1996}{\natexlab{b}}).

\bibitem[{\citenamefont{Yurke and Stoler}(1986)}]{Yurke:1986a}
\bibinfo{author}{\bibfnamefont{B.}~\bibnamefont{Yurke}} \bibnamefont{and}
  \bibinfo{author}{\bibfnamefont{D.}~\bibnamefont{Stoler}},
  \bibinfo{journal}{Phys. Rev. Lett.} \textbf{\bibinfo{volume}{57}},
  \bibinfo{pages}{13} (\bibinfo{year}{1986}).

\bibitem[{\citenamefont{Monroe et~al.}(1996)\citenamefont{Monroe, Meekhof,
  King, and Wineland}}]{Monroe:1996a}
\bibinfo{author}{\bibfnamefont{C.}~\bibnamefont{Monroe}},
  \bibinfo{author}{\bibfnamefont{D.~M.} \bibnamefont{Meekhof}},
  \bibinfo{author}{\bibfnamefont{B.~E.} \bibnamefont{King}}, \bibnamefont{and}
  \bibinfo{author}{\bibfnamefont{D.~J.} \bibnamefont{Wineland}},
  \bibinfo{journal}{Science.} \textbf{\bibinfo{volume}{272}},
  \bibinfo{pages}{1131} (\bibinfo{year}{1996}).

\bibitem[{\citenamefont{O.Scully and Zubairy}(1996)}]{Scully_Quantum:1996a}
\bibinfo{author}{\bibfnamefont{M.}~\bibnamefont{O.Scully}} \bibnamefont{and}
  \bibinfo{author}{\bibfnamefont{M.}~\bibnamefont{Zubairy}},
  \emph{\bibinfo{title}{{Quantum optics}}} (\bibinfo{publisher}{Cambridge.
  University press}, \bibinfo{year}{1996}).

\bibitem[{\citenamefont{H.Press et~al.}(2002)\citenamefont{H.Press,
  A.Teukovsky, T.Vetterling, and P.Flannery}}]{Press_Numerical:2002a}
\bibinfo{author}{\bibfnamefont{W.}~\bibnamefont{H.Press}},
  \bibinfo{author}{\bibfnamefont{S.}~\bibnamefont{A.Teukovsky}},
  \bibinfo{author}{\bibfnamefont{W.}~\bibnamefont{T.Vetterling}},
  \bibnamefont{and}
  \bibinfo{author}{\bibfnamefont{B.}~\bibnamefont{P.Flannery}},
  \emph{\bibinfo{title}{{Numerical recipes in C++}}}
  (\bibinfo{publisher}{Cambridge, University Press, Cambridge},
  \bibinfo{year}{2002}).

\bibitem[{\citenamefont{Narozhny et~al.}(1981)\citenamefont{Narozhny,
  Sanchez-Mondragon, and Eberly}}]{Narozhny:1981a}
\bibinfo{author}{\bibfnamefont{N.~B.} \bibnamefont{Narozhny}},
  \bibinfo{author}{\bibfnamefont{J.~J.} \bibnamefont{Sanchez-Mondragon}},
  \bibnamefont{and} \bibinfo{author}{\bibfnamefont{J.~H.}
  \bibnamefont{Eberly}}, \bibinfo{journal}{Phys. Rev. A.}
  \textbf{\bibinfo{volume}{23}}, \bibinfo{pages}{236} (\bibinfo{year}{1981}).

\bibitem[{\citenamefont{Gea-Banacloche}(1990)}]{JulioGea-Banacloche:1990a}
\bibinfo{author}{\bibfnamefont{J.}~\bibnamefont{Gea-Banacloche}},
  \bibinfo{journal}{Phys. Rev. Lett.} \textbf{\bibinfo{volume}{65}},
  \bibinfo{pages}{3385} (\bibinfo{year}{1990}).

\bibitem[{\citenamefont{Phoenix and Knight}(1991)}]{Phoenix:1991a}
\bibinfo{author}{\bibfnamefont{S.~J.~D.} \bibnamefont{Phoenix}}
  \bibnamefont{and} \bibinfo{author}{\bibfnamefont{P.~L.}
  \bibnamefont{Knight}}, \bibinfo{journal}{Phys. Rev. A.}
  \textbf{\bibinfo{volume}{44}}, \bibinfo{pages}{6023} (\bibinfo{year}{1991}).

\bibitem[{\citenamefont{Gerry and
  Knight}(2004)}]{ChristopherGerry_Introductory:2004a}
\bibinfo{author}{\bibfnamefont{C.}~\bibnamefont{Gerry}} \bibnamefont{and}
  \bibinfo{author}{\bibfnamefont{P.}~\bibnamefont{Knight}},
  \emph{\bibinfo{title}{{Introductory Quantum Optics}}}
  (\bibinfo{publisher}{Cambridge University Press}, \bibinfo{year}{2004}).

\end{thebibliography}
